\documentclass[aps,pre, onecolumn,showpacs, 12pt, showkeys, groupedaddress,amsmath,amssymb]{revtex4-1}
\usepackage{graphicx}  
\usepackage{dcolumn}   
\usepackage{bm}        
\usepackage{verbatim}   

\begin{document}

\title{Ion size effect on electrostatic and electroosmotic properties in soft nanochannels with pH-dependent charge density}
\author{Jun-Sik Sin}
 \email{js.sin@ryongnamsan.edu.kp}
 \affiliation{
   Department of Physics,\textbf {Kim Il Sung} University,Pyongyang, Democratic People's Republic of Korea 
   }
\affiliation{
  Natural Science Center,\textbf {Kim Il Sung} University,
  Pyongyang, Democratic People's Republic of Korea 
   }
\author{Un-Hyok Kim}
\affiliation{Institute of Environmental Science and Water Technology, Academy of Sciences, Pyongyang, Democratic People's Republic of Korea}

\begin{abstract}
We report a theoretical study of  ion size effect on various properties in a soft nanochannel with pH-dependent charge density. We develop a free energy based mean-field theory taking into account ion size as well as pH-dependence of charged polyelectrolyte layer grafted on a rigid surface in an electrolyte. The influence of ion size on properties in a soft nanochannel is evaluated by numerically calculating ion number densities and electrostatic potential. We demonstrate that unlike in point-like ions, for finite sizes of ions, a uniform distribution of chargeable sites within the polyelectrolyte layer causes unphysical discontinuities in ion number densities not only for hydrogen ion but also for other kinds of ions. It is shown that the same cubic spatial distribution of chargeable sites as for point-like ions is necessary to ensure continuity of ion number density and zero ion transport at the polyelectrolyte layer - rigid solid interface. We find that considering finite ion size causes an increase in electrostatic potential and electroosmotic velocity and a decrease in ion number densities. More importantly, we demonstrate that in polyelectrolyte layer, pH-dependence of polyelectrolyte charge density makes accumulation of hydrogen ions stronger than for the other positive ion species in the electrolyte and such a tendency  is further enhanced by considering finite ion size. In addition, we discuss how consideration of finite ion size affects the role of various parameters on electrostatic and electroosmotic properties.
 \end{abstract}


\maketitle
\section{Introduction}
For the last few decades, many researchers have focused great attention on study of polyelectrolyte-grafted interface (charged soft interface). Because charged soft interfaces have been used to deal with electrochemomechanical energy conversion in polymer-grafted nanochannels  (soft nanochannels) \cite%
{Das_SM_2014}, double layer capacitors based on electrical charge storage at a charged soft-interface \cite%
{Das_CSA_2014}, electrokinetic properties of cells, bacteria and viruses \cite%
{Ohshima_Bpc_2005, Ohshima_CSB_1995, Ohshima_JCIS_2003, Ohshima_Bpc_1995,Busscher_Micro_2001,Sasaki_CSB_2003, Modak_CSB_2009, Nguyen_SM_2011, Phan_JCP_2013}, bacterial adhesion to surfaces \cite%
{Busscher_SSR_2002, Elimelech_Lang_2005, Hirata_JCIS_2004}, agglutination of red blood cells \cite%
{Ohshima_VS_2008, Tachev_CSB_2004},  properties of charged soft gels and films \cite%
{Dukhin_JCIS_2005, Dukhin_JCIS_2008, Duval_JCIS_2011,Dukhin_COCIS_2013}, and many more.
	A charged soft interface is constituted by a charged polyelectrolyte layer grafted on a rigid wall in an electrolyte solution \cite%
{Ohshima_ACIS_1995, Duval_PCCP_2011, Ohshima_STAM_2011}. Dissociating chargeable sites on a polyelectrolyte layer allows the polyelectrolyte layer to be charged, producing polyelectrolyte ions. As a consequence, the polyelectrolyte layer - electrolyte interface can play the role of a semi-permeable membrane which only some species of electrolyte ions pass through \cite%
{Das_PRE_2014, Sin_CSA_2017}.
	Many studies focused on the electrostatics of equilibrium electric double layer description combined with the conformation of the polyelectrolyte brushes. On the other hand, many researchers also treated electrokinetics in soft nanochannels including ion transport, electroosmotic transport and electrophoretic mobility of charged soft particles. 
Although different approaches such as Self-Consistent Field theory \cite%
{Taglia_JACS_2010, Egorov_SM_2011, Borisov_JCP_1997, Borisov_Lang_2007, Miklavic_JPC_1988, Zhulina_Macromol_1994}, Mean Field Lattice Theory \cite%
{Linse_EPJE_2001}, Density Functional Theory \cite%
{Pizio_JCP_2013} and Molecular Dynamics Simulation \cite%
{Wu_Macromol_2010, Dobrynin_Lang_2007, Dobrynin_Lang_2011, Dobrynin_Lang_2009, Zhe_Nano_2009, Yan_Biomicro_2011, Binder_JCP_2007} dealt with the above issues, only free energy based mean-field approaches \cite%
{Kirby_SM_2012, Yezek_Lang_2005, Dukhin_JCIS_2004, Duval_Lang_2005, Qian_JPCC_2014, Das_RSC_2015, Das_CSB_2015, Das_JAP_2015,  Das_MN_2016} explicitly considered hydrogen ion number density.
Most studies based on free-energy approaches accounting for pH-dependent polyelectrolyte charge density assume that hydrogen ion number density inside the polyelectrolyte layer are taken as the Boltzmann distribution \cite%
{Busscher_SSR_2002, Kirby_SM_2012, Yezek_Lang_2005, Dukhin_JCIS_2004, Duval_Lang_2005, Qian_JPCC_2014}. i.e. They disregard non-uniformity of hydrogen ion number density distribution due to the chemical reaction which produces the polyelectrolyte layer ions. 

Recently, the authors of \cite%
{Das_RSC_2015, Das_CSB_2015, Das_JAP_2015, Das_MN_2016} pioneered a free energy based mean-field model considering the variation of hydrogen ion number density due to the chemical reaction which produces  the polyelectrolyte layer ions. 
	More importantly, the study of \cite%
{Das_RSC_2015} presents the monomer distribution within the polyelectrolyte layer to obey a non-unique, cubic profile. They also studied that pH affects the electrostatics of a soft spherical particle with a charged core which needs to interpret the electrokinetic properties of biological moieties like MS2 bacteriophage virus \cite%
{Das_CSB_2015}. It is worthwhile to point out that the electroosmotic transport in a soft nanochannel for pH-dependent case is substantially weaker than that for pH-independent charge density  \cite%
{Das_JAP_2015}. However, the studies that consider only point-like ions \cite%
{Das_RSC_2015, Das_CSB_2015, Das_JAP_2015, Das_MN_2016}, disregarded ion size effect in the electrostatics of charged soft surfaces. 

	On the other hand, the significance of ion size effect was already confirmed for a charged soft surface with pH-independent charge density \cite%
{Das_PRE_2014, Sin_CSA_2017} as well as for a hard charged surface \cite%
{Bikerman_PhilosMag_1942, Wicke_ZEC_1952, Iglic_JPhysF_1996, Andelman_PRL_1997, Chu_Biophys_2007, Kornyshev_JPCB_2007, Biesheuvel_JCIS_2007, Li_PRE_2011, Boschitsch_JCC_2012, Iglic_Bioelechem_2010, Siber_PRE_2013, Sin_EA_2015, Iglic_EA_2015}. It is remarkably noticeable that the key result of \cite%
{Das_PRE_2014} is that increasing ion size leads to an increase in Donnan potential and decrease in ion number densities inside polyelectrolyte layer. However, to the best of our knowledge, in none of studies finite ion size is considered for electric double layer of a charged soft surface with pH-dependent charge density. Although in order to determine internal configuration and electric potential in planar negatively charged
lipid head group region the authors of \cite%
{Iglic_Bioelechem_2016} considered the finite size of the charged groups in the region, they disregarded finite sizes of electrolyte ions.

	In this paper, we present a mean-field theory with a simultaneous consideration of pH-dependent charge density of surface charge layer and ion size effect. We derive mathematical expressions for electrostatic potential distribution and ion number densities in a soft nanochannel with pH-dependent charge density and ion size effect. We demonstrate that when considering finite ion size, the cubic monomer distribution suggested by \cite%
{Das_RSC_2015} can ensure the continuities in the value and in the gradient of all the ion number density distributions at the polyelectrolyte-layer- electrolyte interface, zero hydrogen ion flux at the polyelectrolyte-layer - rigid-solid interface and constancy in the total number of polyelectrolyte chargeable sites.  In addition, we discuss how finite ion size does cause changes in electrostatic potential and electroosmotic velocity, and non-monotonic behavior of other positive ion number density profile. 
\section{Theory}
\begin{figure}
\begin{center}
\includegraphics[width=0.7\textwidth]{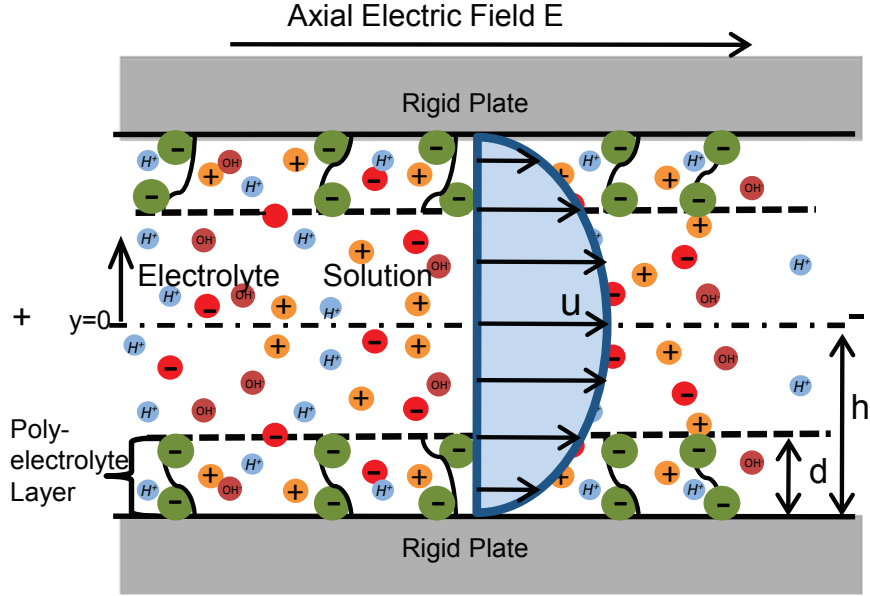}
\caption{(Color online) Schematic illustration of a soft nanochannel. The positive ions, negative ions, hydrogen ions, hydroxyl ions and  polyelectrolyte layer ions are shown in yellow, red, blue, purple and green, respectively.}
\label{fig:1}
\end{center}
\end{figure}

We consider a soft nanochannel of height $2h$, as shown in Fig. \ref{fig:1}. From \cite%
{Das_JAP_2015, Ohshima_STAM_2011, Kirby_SM_2012}, one can know that the polyelecrolyte layer thickness depends mainly on the entropic elastic effect and the excluded volume effect.  Because in the present study we doesn't consider the entropic elastic effect and excluded volume effect of polyelectrolyte layer, we use a constant polyelectrolyte layer thickness for all the calculations. The polyelectrolyte layer has pH-dependent charge and the layer thickness is $d$. It is well-known that the weak acid or base like dissociation of the chargeable site within the polyelectrolyte layer provokes pH-dependence of polyelectrolyte layer charge. Here, we shall study ion size effect on the electrostatic and electrohydrodynamic properties in such a nanochannel.  We shall derive the corresponding electrostatics and use it for the corresponding electroosmotic velocity profile. We take the centerline of the nanochannel as the origin of coordinates and choose the positive y-axis to point top (see Fig. 1).
$\psi$ is the electrostatic potential, while $n_{+}, n_{-}, n_{H^+}$ and $n_{OH^-}$ stand for the number densities of the positive ions, negative ions, hydrogen ions and hydroxyl ions, respectively.
\subsection{Calculation of electrostatic properties with ion size effect}
As in \cite%
{Das_RSC_2015}, we introduce a dimensionless function $\varphi \left( y \right)$ to provide the continuity of  the function $\bar n_{H^+}$ and its derivatives at polyelectrolyte layer - electrolyte interface $(y=-h+d)$. The function shall explicitly represent the spatial dependence of charge distribution within polyelectrolyte layer and satisfy the constraint of the total number of chargeable sites on the polyelectrolyte layer. Considering this number as $N_{p}$ and assuming that $\sigma$  is the area corresponding to a single polyelectrolyte chain and $a$ is the chain thickness, 

\begin{eqnarray}
	\frac{\sigma }{{a^3 }}\int_{ - h}^{ - h + d} {\varphi \left( y \right)dy}  = N_p. 
\label{eq:1}
\end{eqnarray}
When like in \cite%
{Das_RSC_2015}, we take $N_p=\frac{\sigma d}{a^3}$, $\varphi\left(y\right)$ for uniform distribution of chargeable sites should be unity. 
The free energy functional of the present physical system is
\begin{equation}
F = \int {f\left[ {\psi ,n_ +, n_-  , n_{H^ +  } ,n_{OH^ +  } } \right]d^3 {\bf{r}}},
\label{eq:112}
\end{equation}
where accounting for ion size effect and pH-dependent charged density of the polyelectrolyte layer, the free energy density $f$ can be expressed as
\begin{equation}
 \begin{array}{l}
	 f =  - \frac{{\varepsilon _0 \varepsilon _r }}{2}\left| {\nabla \psi } \right|^2  + e_0\psi \left( {n_ +   - n_ -  } \right) + e_0\psi \left( {n_{H^ +  }  - n_{OH^ -  } } \right) - e_0\varphi \left( y \right)n_{A^ -  } \psi  - Ts -\\
 \mu _ +  n_ +   - \mu _ -  n_ -   - \mu _{H^ +  } n_{H^ +  }  - \mu _{OH^ -  } n_{OH^ -  },        - h \le y \le  - h + d,\\
f =  - \frac{{\varepsilon _0 \varepsilon _r }}{2}\left| {\nabla \psi } \right|^2  + e_0\psi \left( {n_ +   - n_ -  } \right) + e_0\psi \left( {n_{H^ +  }  - n_{OH^ -  } } \right) - Ts - \mu _ +  n_ +   - \mu _ -  n_ -   - \mu _{H^ +  } n_{H^ +  }  - \mu _{OH^ -  } n_{OH^ -  }  \\   - h + d \le y \le 0.
 \end{array}
\label{eq:2}
\end{equation}
In Eq. (\ref{eq:2}), $\varepsilon_0$ is the permittivity of free space, $e_0$ is the electronic charge, and $T$ is the temperature.  $\varepsilon_r$ is the relative permittivity of the medium. We assume that the dielectric permittivity in the polyelectrolyte layer is the same as the permittivity in the outside of the layer and thus no image charge effects are present.
Here  the entropy density is $s = k_B \ln W$,  $k_B$ is the Boltzmann constant,  $W$ is the number of configurations, and $n_{A^ - }$  is the number density of polyelectrolyte layer ions ($A^-$). $\mu_{+}, \mu_{-}, \mu_{H^+}$ and $\mu_{OH^-}$ mean the chemical potentials of positive ions,  negative ions, hydrogen ions and hydroxyl ions, respectively.

In order to determine the physical quantities in equilibrium taking into account Eq. (\ref{eq:1}), we should use the method of undetermined multipliers. 

The Lagrangian of the present physical system is
\begin{equation}
 \begin{array}{l}
	L=F- \alpha \left[ {N_p  - \frac{\sigma }{{a^3 }}\int_{ - h}^{ - h + d} {\varphi \left( y \right)dy} } \right],
  \end{array}
\label{eq:215}
\end{equation}
where $\alpha$  is a multiplier of Lagrange. After establishing the Lagrangian the Euler-Lagrange equations can be solved with respect to the ionic number densities and electrostatic potential. Euler-Lagrange equation for an ionic number density can yield mathematical expression of the ion number density in the electrolyte solution, while the Euler-Lagrange equation for electrostatic potential is reduced to Poisson equation. The equation for $\varphi$ should be used to determine charge distribution of polyelectrolyte layer.
We imagine that the polyelectrolyte layer ions are formed by dissociating an acid HA(producing $A^-$ ions, which are the polyelectrolye layer ions), and therefore as obtained in \cite%
{Das_CSA_2014}, we take
\begin{equation}
n_{A^-}=\frac{{\gamma K'_a }}{{\left( {K'_a  + n_{H^ +  } } \right) }},
\label{eq:3}
\end{equation}
where $\gamma$ is the maximum density of the polyelectrolyte chargeable sites and $K'_a =10^3 N_A K_a$($N_A$ is the Avogadro number and $K_a$  is the ionization constant of the acid HA and has units of moles per liter).
We now consider a small volume of the electrolyte, $\Delta V$, in which the ionic number densities $n_{i}, \left(i=+,-,H^+,OH^-\right)$ and electrostatic potential $\psi$ are nearly constant.

The number of configurations representing all ions in the volume $\Delta V$ can be expressed as follows 
\begin{equation}
W = \frac{\left(N \Delta V\right)!}{\left(n_ + \Delta V\right)! \left(n_ - \Delta V\right) !\left(n_{H^ + }\Delta V\right) !\left(n_{OH^ -  }\Delta V\right)!\left(\left(N - n_ +   - n_ -   - n_{H^ +}  - n_{OH^ - }\right) \Delta V\right)!},
\label{eq:4}
\end{equation}
where $N=\frac{1}{v_0}$ and $v_0$ is the volume of an electrolyte ion. i.e. we assume that hydrogen, hydroxyl, negative and positive ions have the same size.

In order to obtain the entropy, $S = k_B \ln W$, we expand the logarithms of factorials using Stirling's formula:
\begin{equation}
\begin{array}{l}
S = k_B \ln W = k_B\left( N\ln N - n_ +  \ln n_ +   - n_ -  \ln n_ -   - n_{H^ +  } \ln n_{H^ +  }  - n_{OH^ -  } \ln n_{OH^ -  }\right) \Delta V- \\  k_B \left(\left( N - n_ +   - n_ -   - n_{H^ +  }  - n_{OH^ -  }  \right)\ln \left(N - n_ +   - n_ -   - n_{H^ +  }  - n_{OH^ -  }\right) \right)\Delta V.
\end{array}
\label{eq:5}
\end{equation}
Here the flexibility of the polyelectrolyte brushes isn't considered and thus the entropy is only related to ionic size and ionic concentration. In particular, it is worthwhile to mention that in the present paper, ionic size effect stems only from the entropy term.
Therefore the entropy density, $s$,  is expressed as follows:
\begin{equation}
\begin{array}{l}
s=\frac{S}{\Delta V}= k_B \left( N\ln N - n_ +  \ln n_ +   - n_ -  \ln n_ -   - n_{H^ +  } \ln n_{H^ +  }  - n_{OH^ -  } \ln n_{OH^ -  }\right)- \\ k_B \left( {N - n_ +   - n_ -   - n_{H^ +  }  - n_{OH^ -  } } \right)\ln \left( {N - n_ +   - n_ -   - n_{H^ +  }  - n_{OH^ -  } } \right).
\end{array}
\label{eq:5}
\end{equation}

Minimizing Eq. (\ref{eq:215}) with respect to the variables $\psi, n_{\pm}, n_{H^+}$ and $n_{{OH}^-}$ yields governing equations (see Appendix for detailed derivation).

For feasible calculation of the present problem, the dimensionless parameters are introduced by
\begin{equation}
\begin{array}{l}
n_b=n_{+ ,\infty}=n_{- ,\infty}, \bar n_ +   = \bar n_ +  /n_b ,\bar n_ -   = \bar n_ -  /n_b , \bar n_{H^ +  }  = n_{H^ +  } /n_b, \bar n_{OH^ -  }  = n_{OH^ -  } /n_b, \bar K'_a=\bar K'_a /n_b, \\
\bar \gamma  = \gamma /n_b, \bar \psi  = e_0 \psi/\left(k_B T\right),\bar y = y/h, \bar \lambda  = \lambda /h, \lambda  = \sqrt {\frac{{\varepsilon _0 \varepsilon _r k_B T}}{{2e_0^2 n_b }}}, \frac{d}{{dy}} = \frac{d}{{d\bar y}}\frac{1}{h}.
\end{array}
\label{eq:31}
\end{equation}
In the region of $ -1  \le \bar y \le -1 + \bar d $, we obtain the following equation for determining the electrostatic potential 
\begin{equation}
\frac{{d^2 \bar \psi }}{{d\bar y^2 }} = \frac{1}{{\bar \lambda ^2 }}\left( { - \left( {\bar n_ +   - \bar n_ -  } \right) - \left( {\bar n_{H^ +  }  - \bar n_{OH^ -  } } \right) + \varphi \left( y \right)\frac{{\bar K'_a \bar \gamma }}{{\left( {\bar K'_a  + \bar n_{H^ +  } } \right)}}} \right),
\label{eq:30}
\end{equation}
In the region (i.e. inside the polyelectrolyte layer), the number densities of ions are the below equations Eqs. (\ref{eq:33}, \ref{eq:34}, \ref{eq:35}, \ref{eq:36}), which are  different from previous one \cite%
{Das_RSC_2015}.
\begin{equation}
\small
\bar n_{H^ +  }  = \frac{{\bar n_{H^ +  ,\infty } \exp \left( { - \bar \psi \left( {1 + \varphi \left( y \right)\frac{{\bar \gamma \bar K'_a }}{{\left( {\bar K'_a  + \bar n_{H^ +  } } \right)^2 }}} \right)} \right)}}{{\left( {1 - 2\nu  -  \bar n_{H^ +  ,\infty } \nu } -  \bar n_{OH^ -  ,\infty }\nu \right) + 2\nu \cosh \left( {\bar \psi } \right) +\bar n_{OH^ - ,\infty  } \nu \exp \left( {\bar \psi } \right) + \bar n_{H^ +  ,\infty } \nu  \exp \left( { - \bar \psi \left( {1 + \varphi \left( y \right)\frac{{\bar \gamma \bar K'_a }}{{\left( {\bar K'_a  + \bar n_{H^ +  } } \right)^2 }}} \right)} \right)}},
\label{eq:33}
\end{equation}
\begin{equation}
\small
\bar n_ -   = \frac{{\exp \left( {\bar \psi } \right)}}{{\left( {1 - 2\nu  -  \bar n_{H^ +  ,\infty } \nu } -  \bar n_{OH^ -  ,\infty }\nu \right) +  2\nu \cosh \left( {\bar \psi } \right) +\bar n_{OH^ - ,\infty  } \nu \exp \left( {\bar \psi } \right) + \bar n_{H^ +  ,\infty } \nu  \exp \left( { - \bar \psi \left( {1 + \varphi \left( y \right)\frac{{\bar \gamma \bar K'_a }}{{\left( {\bar K'_a  + \bar n_{H^ +  } } \right)^2 }}} \right)} \right)}},
\label{eq:34}
\end{equation}
\begin{equation}
\small
\bar n_ +   = \frac{{\exp \left( { - \bar \psi } \right)}}{{\left( {1 - 2\nu  -  \bar n_{H^ +  ,\infty } \nu } -  \bar n_{OH^ -  ,\infty }\nu \right) + 2\nu \cosh \left( {\bar \psi } \right) +\bar n_{OH^ - ,\infty  } \nu \exp \left( {\bar \psi } \right) + \bar n_{H^ +  ,\infty } \nu  \exp \left( { - \bar \psi \left( {1 + \varphi \left( y \right)\frac{{\bar \gamma \bar K'_a }}{{\left( {\bar K'_a  + \bar n_{H^ +  } } \right)^2 }}} \right)} \right)}},
\label{eq:35}
\end{equation}
\begin{equation}
\small
\bar n_{OH^ -  }  = \frac{{\bar n_{OH^ -  ,\infty } \exp \left( {\bar \psi } \right)}}{{\left( {1 - 2\nu  -  \bar n_{H^ +  ,\infty } \nu } -  \bar n_{OH^ -  ,\infty }\nu \right) + 2\nu \cosh \left( {\bar \psi } \right) +\bar n_{OH^ - ,\infty  } \nu \exp \left( {\bar \psi } \right) + \bar n_{H^ +  ,\infty } \nu  \exp \left( { - \bar \psi \left( {1 + \varphi \left( y \right)\frac{{\bar \gamma \bar K'_a }}{{\left( {\bar K'_a  + \bar n_{H^ +  } } \right)^2 }}} \right)} \right)}},
\label{eq:36}
\end{equation}
where $\bar n_{H^+ ,\infty}=n_{H^+ ,\infty}/n_b$, $\bar n_{OH^- ,\infty}=n_{OH^- ,\infty}/n_b$ and $\nu=n_b v_0$ is the steric factor.

In the region of  $ -1 + \bar d \le \bar y \le 0$, we obtain
\begin{equation}
\frac{{d^2 \bar \psi }}{{d\bar y^2 }} = \frac{1}{{\bar \lambda ^2 }}\left( { - \left( {\bar n_ +   - \bar n_ -  } \right) - \left( {\bar n_{H^ +  }  - \bar n_{OH^ -   } } \right)} \right),
\label{eq:37}
\end{equation}
\begin{equation}
\bar n_{H^ +  }  = \frac{{\bar n_{H^ +  ,\infty } \exp \left( { - \bar \psi } \right)}}{{\left( {1 - 2\nu  - \bar n_{H^ +  ,\infty } \nu - \bar n_{OH^ -  ,\infty } \nu} \right) +2\nu \cosh \left( { \bar \psi } \right) + \bar n_{OH^ -  ,\infty} \nu \exp \left( {\bar \psi } \right) + \bar n_{H^ +  ,\infty } \nu \exp \left( { - \bar \psi } \right)}},
\label{eq:38}
\end{equation}
\begin{equation}
\bar n_ -   = \frac{{\exp \left( {\bar \psi } \right)}}{{\left( {1 - 2\nu  - \bar n_{H^ +  ,\infty } \nu - \bar n_{OH^ -  ,\infty } \nu} \right) +2\nu \cosh \left( { \bar \psi } \right) + \bar n_{OH^ -  ,\infty} \nu \exp \left( {\bar \psi } \right) + \bar n_{H^ +  ,\infty } \nu \exp \left( { - \bar \psi } \right)}},
\label{eq:39}
\end{equation}
\begin{equation}
\bar n_ +   = \frac{{\exp \left( { - \bar \psi } \right)}}{{\left( {1 - 2\nu  - \bar n_{H^ +  ,\infty } \nu - \bar n_{OH^ -  ,\infty } \nu} \right) +2\nu \cosh \left( { \bar \psi } \right) + \bar n_{OH^ -  ,\infty} \nu \exp \left( {\bar \psi } \right) + \bar n_{H^ +  ,\infty } \nu \exp \left( { - \bar \psi } \right)}},
\label{eq:40}
\end{equation}
\begin{equation}
\bar n_{OH^ -  }  = \frac{{\bar n_{OH^ -  ,\infty } \exp \left( {\bar \psi } \right)}}{{\left( {1 - 2\nu  - \bar n_{H^ +  ,\infty } \nu - \bar n_{OH^ -  ,\infty } \nu} \right) +2\nu \cosh \left( { \bar \psi } \right) + \bar n_{OH^ -  ,\infty} \nu \exp \left( {\bar \psi } \right) + \bar n_{H^ +  ,\infty } \nu \exp \left( { - \bar \psi } \right)}}.
\label{eq:41}
\end{equation}
The coupled set of Eqs. (\ref{eq:28}-\ref{eq:41}) should be solved by combining with the following boundary conditions
\begin{equation}
\begin{array}{l}
 \left( {\frac{{d\bar \psi}}{{d\bar y}}} \right)_{\bar y =  - 1}  = 0;\left( {\frac{{d\bar \psi}}{{d\bar y}}} \right)_{\bar y = 0}  = 0; \\ 
 \left( {\frac{{d\bar \psi}}{{d\bar y}}} \right)_{\bar y = \left( { - 1 + \bar d} \right)^ +  }  = \left( {\frac{{d\bar \psi}}{{d\bar y}}} \right)_{\bar y = \left( { - 1 + \bar d} \right)^ -  } ; \\ 
 \left( {\bar \psi} \right)_{\bar y = \left( { - 1 + \bar d} \right)^ +  }  = \left( {\bar \psi} \right)_{\bar y = \left( { - 1 + \bar d} \right)^ -  } . \\ 
 \end{array}
\label{eq:412}
\end{equation}
	If we neglect ion size, i.e. $\nu=0$, our equations are reduced to those when considering only pH-dependent charge within polyelectrolyte layer and disregarding ion size effect \cite%
{Das_RSC_2015, Das_CSB_2015, Das_JAP_2015, Das_MN_2016}.
\subsection{Calculation of electroosmotic velocity with ion size effect}
In an externally applied axial electric field $E$, we study a steady-state, fully developed axial electroosmotic transport in a soft nanochannel with pH-dependent charge density. 

The equation for the velocity field u should be the same as in \cite%
{Das_JAP_2015, Das_JCIS_2015},
\begin{equation}
\begin{array}{l}
 \eta \frac{{d^2 u}}{{dy^2 }} + e_0\left( {n_ +   - n_ -   + n_{H^ +  }  - n_{OH^ -  } } \right)E - \mu _c u = 0, - h \le y \le  - h + d, \\ 
 \eta \frac{{d^2 u}}{{dy^2 }} + e_0\left( {n_ +   - n_ -   + n_{H^ +  }  - n_{OH^ -  } } \right)E = 0,  - h + d \le y \le 0,\\ 
\end{array}
\label{eq:42}
\end{equation}
where $\eta$ is the dynamic viscosity of the liquid, $\mu _c  = \left( {\frac{{\varphi \left( y \right)}}{b}} \right)^2 $  is the drag coefficient within the polyelectrolyte layer and $b$ is the effective monomer size. As in Ref. [46], we can express $\mu _c  = \left( {\frac{{\varphi \left( y \right)}}{b}} \right)^2$. Transforming the above equations into the dimensionless form gives the following equation;
\begin{equation}
\begin{array}{l}
\frac{{d^2 \bar u}}{{d\bar y^2 }} + \frac{{\bar E}}{{{\bar \lambda} ^2 }}\left( {\bar n_ +   - \bar n_ -   + \bar n_{H^ +  }  -\bar n_{OH^ -  } } \right) - \bar \alpha ^2 \varphi ^2 \left( y \right)\bar u = 0, - 1  \le \bar y \le - 1 + \bar d\\
\frac{{d^2 \bar u}}{{d\bar y^2 }} + \frac{{\bar E}}{{{\bar \lambda} ^2 }}\left( {\bar n_ +   - \bar  n_ -   + \bar n_{H^ +  }  - \bar n_{OH^ -  } } \right) = 0,  - 1 + \bar d \le \bar y \le 0, \\
\end{array}
\label{eq:43}
\end{equation}
where $\bar u=\frac{u}{u_0}, u_0=\frac{k_B T}{e}\frac{\varepsilon_0 \varepsilon_r E_0}{\eta}, \bar E=\frac{E}{E_0}, \bar \alpha=\frac{h}{b \sqrt{\eta}}$ and $E_0$ is       
 the scale of the electric field \cite%
 {Das_JAP_2015}.

The dimensionless electroosmotic velocity $\bar u$ should be calculated by solving Eq. (\ref{eq:43}) in presence of the dimensionless ion number densities and dimensionless electrostatic potential being results from the previous subsection, and the boundary conditions written
in Eq. (\ref{eq:44}).
\begin{equation}
\begin{array}{l}
 \left( {\frac{{d\bar u}}{{d\bar y}}} \right)_{\bar y =  - 1}  = 0, \left( {\frac{{d\bar u}}{{d\bar y}}} \right)_{\bar y = 0}  = 0, \\ 
 \left( {\frac{{d\bar u}}{{d\bar y}}} \right)_{\bar y = \left( { - 1 + \bar d} \right)^ +  }  = \left( {\frac{{d\bar u}}{{d\bar y}}} \right)_{\bar y = \left( { - 1 + \bar d} \right)^ -  } , \\ 
 \left( {\bar u} \right)_{\bar y = \left( { - 1 + \bar d} \right)^ +  }  = \left( {\bar u} \right)_{\bar y = \left( { - 1 + \bar d} \right)^ -  } . \\ 
 \end{array}
\label{eq:44}
\end{equation}

\section{Results and Discussion}
Using the fourth order Runge-Kutta method, we obtain $\bar n_{H^+}$,  $\bar n_{OH^-}$, $\bar n_{+}$, $\bar n_{-}$, $\bar \psi$, $\bar u$  by combining  Eqs. (\ref{eq:25}) - (\ref{eq:43}) with the boundary conditions. In all the calculations, the temperature $T$  is taken as $298K$ , respectively.
\begin{figure}
\includegraphics[width=1\textwidth]{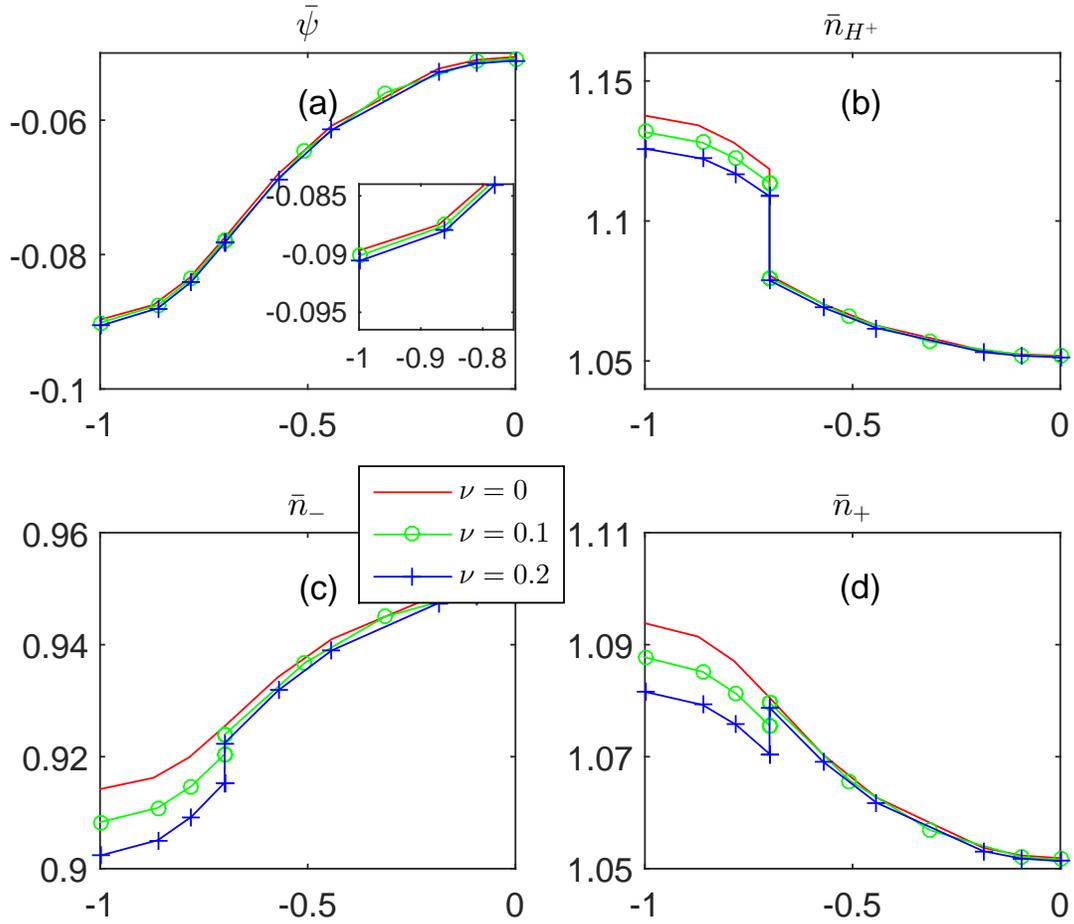}
\caption{(Color online) Transverse variation of the dimensionless electrostatic potential (a), the dimensionless hydrogen ion number density (b), the dimensionless negative ion number density (c) and the dimensionless positive ion number density (d)  for uniform distribution of polyelectrolyte chargeable sites  within the polyelectrolyte layer as functions of $\bar y$-coordinate.  For all the figures we take the five parameters( $K_\lambda  ,\bar \lambda ,\bar \alpha ,\bar n_{H^ +  ,\infty } ,K'_a$)  as unity and $\bar d = 0.3$. }
\label{fig:2}
\end{figure}

Fig.  \ref{fig:2}(a )-(d) display the trasverse variation of the dimensionless electrostatic potential($\bar \psi$) , the variation of the dimensionless hydrogen ion number density ($\bar n_{H^+}$), negative ion number density ($\bar n_{-}$) and positive ion number density ($\bar n_{+}$) profiles as functions of $\bar y$-coordinate.  Here, solid line, circles and crosses represent the cases of $\nu=0,0.1,0.2$, respectively.

Fig. \ref{fig:2}(a) shows the spatial variation of the electrostatic potential. We find that increase in ion size results in an increase in magnitude of electrostatic potential. This is attributed to the fact that the larger the ion size, the weaker the screening property. Such a behavior is consistent with the result of \cite%
{Das_PRE_2014} where pH-dependence is not considered.

Fig. \ref{fig:2}(b) indicates the transverse variation of hydrogen ion number density as a function of $\bar y$-coordinate. For all cases of ion size, at the interface between electrolyte and polyelectrolyte layer there exist a discontinuity of the distribution.  This is easily understood by comparing Eq. (\ref{eq:33}) and Eq. (\ref{eq:38}). This is in agreement with the result of \cite%
{Das_RSC_2015}, where consideration of ion size lowers negative ion and positive ion number densities inside polyelectrolyte layer.
It also shows that an increase in ionic size reduces the hydrogen ion number density. 
In fact, a large size of ions means an enhancement in repulsion between ions.
As a consequence, the enhanced repulsion will diminish accumulation of counterions due to electrostatic attraction near a charged polyelectrolyte layer.

Fig. \ref{fig:2}(c) and Fig. \ref{fig:2}(d) demonstrate the variation of negative and positive ion number densities with $\bar y$-coordinate, respectively. A noticeable fact is that for finite ion sizes, there exists a jump in every ion number density. In fact, for the case of point-like ion, number densities of other ions except for hydrogen ion are continuous functions \cite%
{Das_RSC_2015}.  However, consideration of finite ion size makes the discontinuity of the number densities at the interface for all species of ions. This is proved by comparing Eqs. (\ref{eq:34}, \ref{eq:35}) and Eqs. (\ref{eq:39}, \ref{eq:40}). This is a hitherto unknown result.
 \begin{figure}
 \includegraphics[width=1\textwidth]{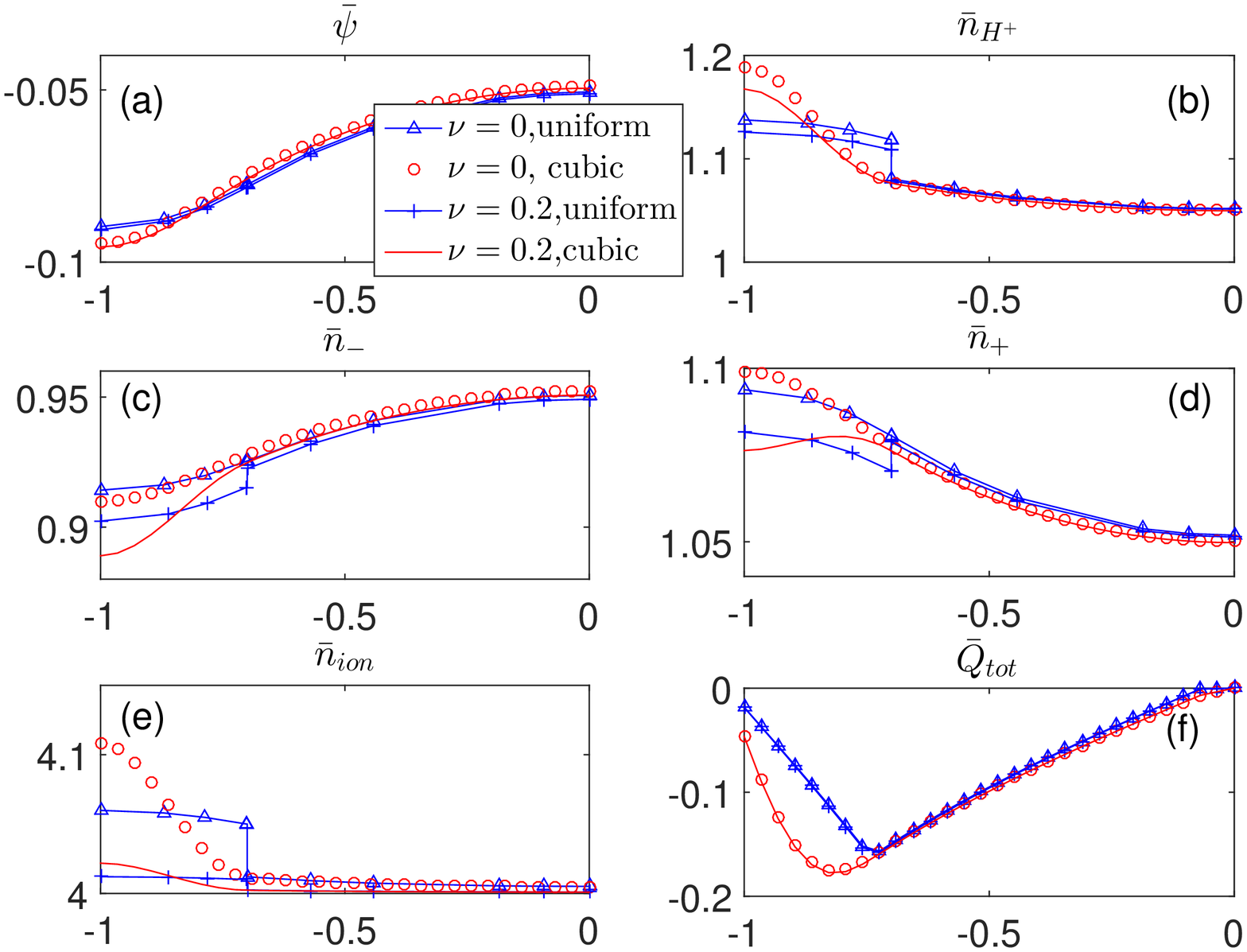}
\caption{(Color online) The dimensionless electrostatic potential (a) , the dimensionless hydrogen ion number density (b), the dimensionless negative ion number density (c), the dimensionless positive ion number density (d), the dimensionless total ion number density (e) and the dimensionless accumulated charge near the nanochannel wall  (f)   as a function of the y- coordinate for both a uniform and the cubic distribution of polyelectrolyte chargeable sites  within the polyelectrolyte layer as functions of $\bar y$-coordinate.  Triangles, circles, crosses and solid line respectively represent the cases of $\nu=0, 0.2$ for a uniform or the cubic monomer distributions. Other parameters are the same as in Fig. \ref{fig:2}.}
\label{fig:3}
\end{figure}
\begin{figure}
\includegraphics[width=0.5\textwidth]{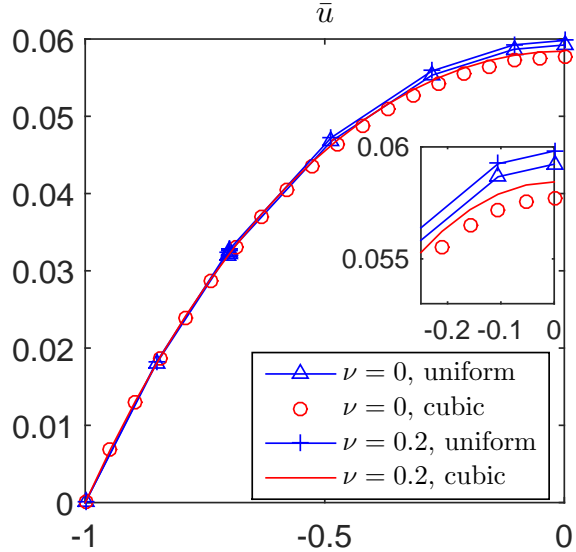}
\caption{(Color online)  The dimensionless electroosmotic velocity as a function of the y- coordinatel for both uniform and the cubic distribution of polyelectrolyte chargeable sites  within the polyelectrolyte layer.  Triangles, circles, crosses and solid line respectively represent the cases of $\nu=0, 0.2$ for a uniform or the cubic monomer distributions. Other parameters are the same as in Fig. \ref{fig:2}.}
\label{fig:4}
\end{figure}
\begin{figure}
\includegraphics[width=1\textwidth]{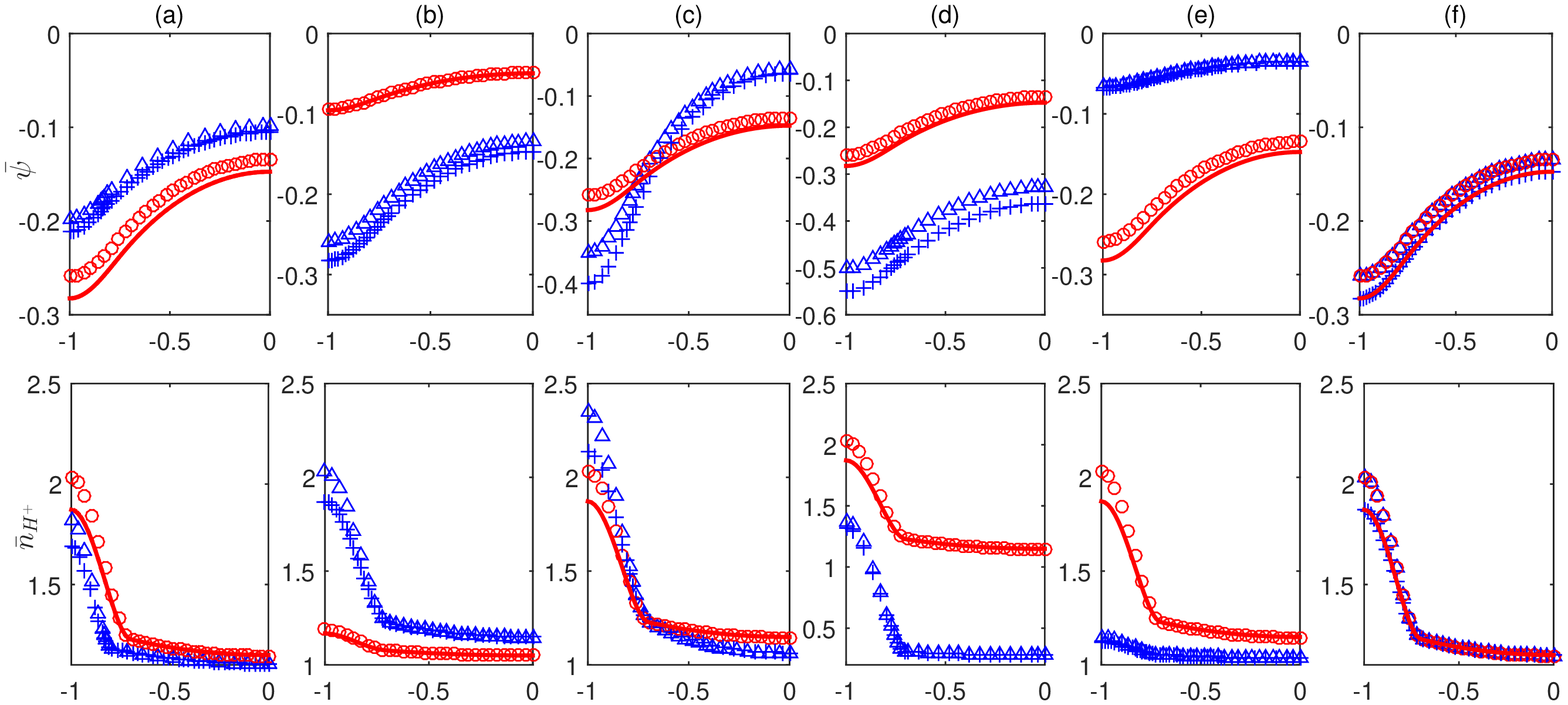}
\caption{(Color online) The dimensionless electrostatic potential (top panel) and the dimensionless hydrogen ion number density(bottom panel) for the cubic distribution of polyelectrolyte chargeable sites  within the polyelectrolyte layer for different parameters  $\bar d, K_\lambda  ,\bar \lambda ,\bar \alpha ,\bar n_{H^ +  ,\infty } ,K'_a$. For all the figures, we take four parameters( $\bar \lambda,\bar n_{H^ +  ,\infty } ,K'_a, ,\bar \alpha $) as unity and $K_\lambda   = 0.5, \bar d = 0.3$, except for (a) $\bar d = 0.2, \nu=0$ (Triangles); $\bar d = 0.3,\nu = 0$
 (Circles);  $\bar d = 0.2, \nu = 0.2$ (Crosses); $\bar d = 0.3, \nu = 0.2$ (Solid line) ; (b) $K_\lambda   = 0.5, \nu=0$ (Triangle); $K_\lambda   = 0.5, \nu=0.2$ (Crosses); $K_\lambda   = 1, \nu=0$(Circles); $K_\lambda   = 1, \nu=0.2$   (Solid line); (c) $\bar \lambda=0.5, \nu=0$ (Triangles); $\bar \lambda=0.5, \nu=0.2$ (Crosses); $\bar \lambda=1, \nu=0$ (Circles); $\bar \lambda=1, \nu=0.2$  (Solid line); (d) $\bar n_{H^ +  ,\infty }  = 0.2, \nu=0$ (Triangle); $\bar n_{H^ +  ,\infty }  = 0.2, \nu=0.2$  (Crosses); $\bar n_{H^ +  ,\infty }  = 1, \nu=0$  (Circles); $\bar n_{H^ +  ,\infty }  = 1, \nu=0.2$ (Solid line); (e) $K'_a  = 0.1, \nu=0$ (Triangles); $K'_a  = 0.1,\nu=0.2$ (Crosses); $K'_a  = 1,\nu=0$ (Circles);  $K'_a=1, \nu=0.2$ (Solid line);  (f) $\bar \alpha  = 0.1, \nu=0$ (Triangles); $\bar \alpha  = 0.1, \nu=0.2$ (Crosses); $\bar \alpha  = 1, \nu=0$ (Circles); $\bar \alpha  = 1, \nu=0.2$ (Solid line) }
\label{fig:5}
\end{figure}
In order to avoid such a unphysical behavior, an appropriate distribution for polyelectrolyte chargeable sites should be presented.

Fig. \ref{fig:3}(a) shows the dimensionless electrostatic potential by using the cubic distribution function Eq. (\ref{eq:27}) for polyelectrolyte chargeable sites. It is worthwhile to note that for both cases of a uniform and the cubic distribution functions, electrostatic potential for ions with finite size is large compared to that for a point-like ion. Although at the interface between the rigid plate and the polyelectrolyte layer $(\bar y=-1)$, the magnitudes of electrostatic potentials for the cubic distribution function are larger than that for a uniform distribution function, at the centerline of nanochannel the order of magnitudes of electrostatic potentials  is inversed. This is understood by the fact that the cubic distribution function increases with decreasing the distance from rigid plate.

Fig. \ref{fig:3}(b) and Fig. \ref{fig:3}(c) display the dimensionless hydrogen ion number density and the dimensionless negative ion number density versus the $\bar y$-coordinate. As shown in Fig. \ref{fig:3}(b), for the cases of $\nu=0, 0.2$ having the cubic distribution function, the hydrogen and negative ion number density profiles have no discontinuities in their domains of definition($-1\le \bar y \le 0$). It should be mentioned that in the similar way in \cite%
{Das_PRE_2014}, the ion number densities for the case of finite ion size($\nu=0.2$) are lower than those for point-like ions($\nu=0$). 

Fig. \ref{fig:3}(d) quantifies the dimensionless positive ion number density profiles for different cases. Noticeably, it is shown that for the case of the cubic distribution function for chargeable sites, consideration of finite ion size makes non-monotonic positive ion number density profile.  It should be noted that positive ion number density first increases with decreasing the distance from the rigid plate and then reaches to a maximum value at the interface between electrolyte and polyelectrolyte layer, again decreases with decreasing the distance from the rigid plate. This is one of key results of this paper. Comparing Eq. (\ref{eq:35}) with Eq. (\ref{eq:33}), we can demonstrate that this fact is due to the preferential accumulation of hydrogen ions inside the polyelectrolyte layer.

Fig.\ref{fig:3}(e) shows total ion distribution profile, i.e, $\bar n_{ion}  = \bar n_ +   + \bar n_ -   + \bar n_{H^ +  }  + \bar n_{OH^ +  }$.
Fig. \ref{fig:3}(e) explicitly demonstrates that the total ion density in polyelectrolyte layer is higher than that in the nanochannel centerline. 
In fact, it's known that  in a neutral slit containing electrolyte, the multipole interaction in the charged system provides the density of the electrolyte higher in the slit center than near the slit boundary.
Although we consider the nanochannel without wall charge, the nanochannel has a charged polyelectrolyte layer which can play the role of charged wall. Therefore, we can't expect such a phenomenon as in a neutral slit. 

Fig. \ref{fig:3}(f) shows accumulated charge near the nanochannel wall according to the distance from the wall, $Q_{tot} \left( {\bar y} \right) = \int_{ - 1}^{\bar y} {\left( {\bar n_ +   - \bar n_ -   + \bar n_{H^ +  }  - \bar n_{OH^ +  }  - \bar n_{A^ -  } } \right)d\bar r}$.

For all the approaches we study here, Fig. \ref{fig:3}(f) clearly displays $Q_{tot} \left( {\bar y = 0} \right) = 0$.
As a consequence, our calculations ensure the important fact that charge of polyelectrolyte layer is completely neutralized by the ions in the electrolyte.

Fig. \ref{fig:4} shows the dimensionless electroosmotic velocity profile for different cases. 
At the centerline of nanochannel, the velocity values for the cubic distribution functions are lower than corresponding ones for uniform distribution. This can be understood as follows. In fact, as mentioned in Fig. \ref{fig:3}(a), at the centerline of the nanochannel the magnitudes of electrostatic potential for the cubic distribution function are lower than corresponding ones for uniform distribution function.  It is well-known that for the given set of parameters, smaller magnitudes of electrostatic potentials result in smaller differences between positive and negative ion number densities. Consequently, combining the above fact with Eq. (\ref{eq:43}), we can easily understand that at the centerline of the nanochannel the velocity values for the cubic monomer distribution are small compared to those for uniform distribution.

Fig. \ref{fig:5}(a)-(f) show the effect of the different parameters on dimensionless electrostatic potential profiles (top panel) and the dimensionless hydrogen ion number density profiles (bottom panel). 
First of all, it should be noted that for all the sets of the parameters, consideration of ion size causes an increase in electrostatic potential and a decrease in hydrogen number density. As can be seen from Fig. \ref{fig:5}(a), such effects are enhanced by an increase in $\bar d$. In fact, as pointed out in \cite%
{Das_JAP_2015}, the magnitude of electrostatic potential increases with an increase in $\bar d$ because the larger is  $\bar d$, the larger becomes amount of polyelectrolyte layer charge. Then it is well known that the larger the amount of polyelectrolyte layer charge, the stronger the variation in electrostatic potential and counterion number density due to finite ion size \cite%
{Das_PRE_2014}. In a similar way, we confirm from Fig. \ref{fig:5}(b) and Fig. \ref{fig:5}(e) that small $K_\lambda$ and large $K'_a$ mean a increased amount of polyelectrolyte layer charge, and consequently produce large variations in electrostatic potential and hydrogen ion number density due to ion size effect. Fig. \ref{fig:5}(c) shows that decrease in $\bar \lambda$ requires an enhanced electrostatic potential at the polyelectrolyte layer-rigid plate interface to attract more hydrogen ions inside polyelectrolyte layer since for the smaller $\bar \lambda$, numbers of hydrogen ions outside polyelectrolyte layer are smaller than corresponding ones for large $\bar \lambda$. Consequently, for the small $\bar \lambda$, consideration of finite ion size results in a larger variation in electrostatic potential and hydrogen ion number density.  Fig. \ref{fig:5}(d) demonstrates that decrease in $\bar n_{H^+,\infty}$ causes increasing total charge of polyelectrolyte layer according to the Eq. (\ref{eq:3}) and finally increasing the electrostatic potential. Therefore, although finite ion size hardly affects hydrogen ion number density, electrostatic potential is affected by the effect. Fig. \ref{fig:5}(f) illustrates that ion size effect is irrelevant to the role of $\bar \alpha$ parameter.

\begin{figure}
\includegraphics[width=1\textwidth]{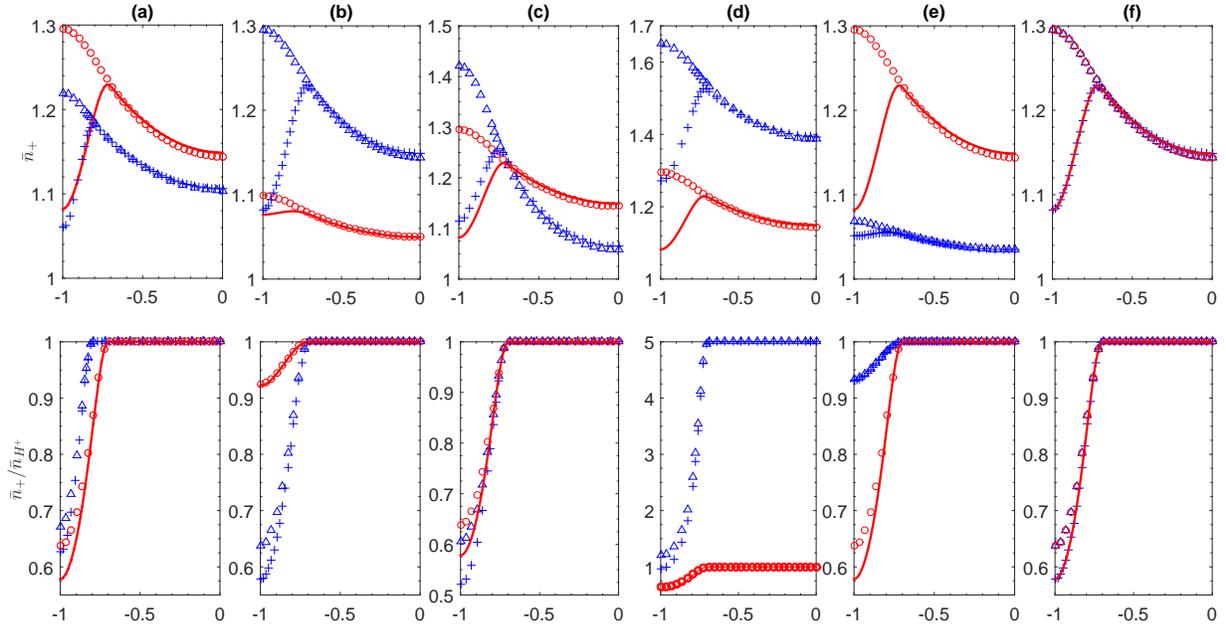}
\caption{(Color online) The dimensionless positive ion number density (top panel) and the ratio of positive ion number density to  hydrogen ion number density(bottom panel)  for the cubic distribution of polyelectrolyte chargeable sites  within the polyelectrolyte layer for different parameters $\bar d,K_\lambda  ,\bar \lambda ,\bar n_{H^ +  ,\infty } , K'_a, \bar \alpha$. All the parameters for all the figures are the same as in Fig. \ref{fig:5}.}
\label{fig:6}
\end{figure}

Fig. \ref{fig:6}(a)-(f) show positive ion number density profiles (top panel) and ratios of hydrogen ion numbers to positive ion numbers (bottom panel) for different cases. For all the cases of Fig. \ref{fig:6}(a)-(f), we identify that increasing ion size lowers positive ion number density.  To quantify such depletion of positive ions,  we introduce the ratio of positive ion number density to hydrogen ion number density as a function of the y-coordinate (bottom pannel). For all the cases of Fig. \ref{fig:6}(a)-(f), it is shown that when ion size increases, the ratio of positive ion number density to hydrogen ion number density inside polyelectrolyte layer significantly decreases. It can be confirmed from the fact that consideration of finite ion size diminishes the ratio of positive ion number density to hydrogen ion number density according to Eqs. (\ref{eq:33})-(\ref{eq:35}) since ion size effect increases electrostatic potential.
\begin{figure}
\includegraphics[width=1\textwidth]{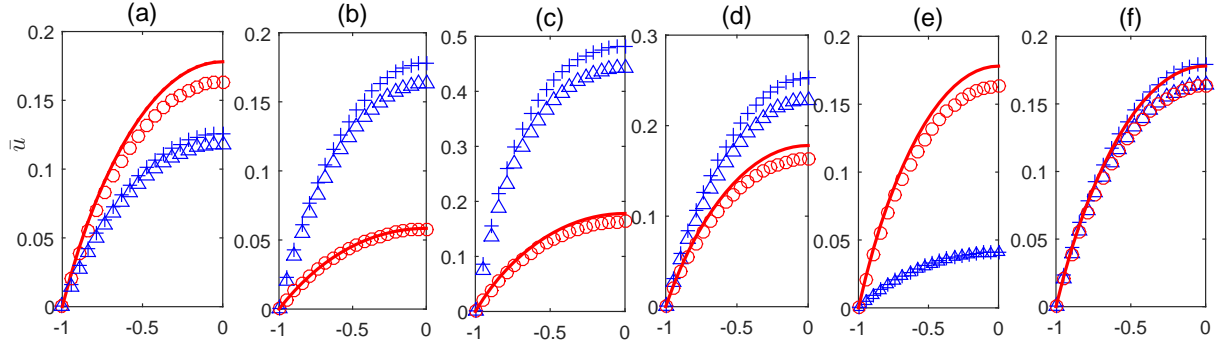}
\caption{(Color online) The dimensionless electroosmotic velocity for the cubic distribution of polyelectrolyte chargeable sites  within the polyelectrolyte layer for different parameters $\bar d,K_\lambda, \bar \lambda ,\bar n_{H^ +  ,\infty }, K'_a, \bar \alpha$.  All the parameters for all the figures are the same as in Fig. \ref{fig:5}. }
\label{fig:7}
\end{figure}
As can be seen in Fig. \ref{fig:6}(a), the ratio outside polyelectrolyte layer is one, which is understood from Eq. (31) and Eq. (29). However, inside polyelectrolyte layer the ratio is decreased with decreasing the distance from the rigid-plate.  Fig. \ref{fig:6}(a) clearly illustrates that a long polyelectrolyte layer and a large ion size cause a stronger depletion of positive ions. Comparing Eq. (\ref{eq:33}) and Eq. (\ref{eq:35}), we prove the fact which increasing hydrogen ion number density makes the decrease in the ratio. Increase in ion size and  $\bar d$ decreases the ratio. 

Fig. \ref{fig:6}(b) shows that the ratio of positive ion numbers to hydrogen ion numbers is decreased by increasing $K_\lambda$ due to lowering the electrostatic potential. In the same way, we can see that the ratio increases with increasing $K'_a$ [see Fig. \ref{fig:6}(c)] and decreasing $\bar \lambda$[see Fig. \ref{fig:6}(e)]. Fig. \ref{fig:6}(d) represents that small $\bar n_{H^+,\infty}$ provides a large variation in the ratio due to the preferential accumulation of hydrogen ions. 

Fig. \ref{fig:7}(a)-(f) show the effect of different parameters on the dimensionless electroosmotic velocity.
Fig. \ref{fig:7}(a)-(f) exhibit that increasing ion size makes electroosmotic velocity to increase. This is attributed to the increase in electrostatic potential due to ion size effect.
Fig. \ref{fig:7}(a)-(e) also quantifies that increasing $\bar d$(see Fig. \ref{fig:7}(a))and $K'_a$(see Fig. \ref{fig:7}(e)) and decreasing $K_\lambda$(see Fig. \ref{fig:7}(b)),  $\bar \lambda$(see Fig. \ref{fig:7}(c)) and  $\bar n_{H^+,\infty}$(see Fig. \ref{fig:7}(d)) enhance the variation in electroosmotic velocity due to ion size effect.

In the present paper, we didn't consider the dielectric discontinuity effects, the flexibility of the polyelectrolyte chains, and the multipole interactions in the system.
In the future, we will further study the influence of such factors.
To treat such systems, we can use not only our approach but also  Molecular Dynamics Simulation \cite%
{Qiao_2004} and Classical Fluid Density Functional Theory approaches \cite%
{Lee_2012}.

\section{Conclusions}
In this work, we provide a free-energy based mean-field theory taking into account ion size effect to describe electrostatics of a charged soft surface with pH-dependent charge density. First of all, we show that consideration of ion size yields that uniform polyelectrolyte chargeable sites distribution within polyelectrolyte layer produce unphysical discontinuities for all types of ions. We prove that the cubic distribution condition is able to properly remove the jumps and ensure the physical conditions. It is illustrated that finite ion size causes an increase in electrostatic potential and electroosmotic velocity, and non-monotonic behavior of the number density profile for positive ion other than hydrogen ion in an electrolyte solution. Moreover, it turns out that ion size effect triggers enhanced preferential accumulation of hydrogen ions inside polyelectrolyte layer compared to other positive ion. Finally, it is also demonstrated that ion size effect enhances the effect of different parameters on electrostatic and electroosmotic properties in soft nanochannels. We reach the conclusion that simultaneous consideration of ion size effect and pH-dependent charge density is mandatory for well studying charged soft surfaces and soft nanochannels.    

\section{Appendix: Derivation of governing equation}

We employ a variational approach to derive electrostatic potential and number density of ions in an electrolyte.

Minimizing Eq. (\ref {eq:215}) with respect to $\psi ,n_ +  ,n_ -  ,n_{H^ +  } ,n_{OH^ -  } ,\varphi \left( y \right)$,  we arrive at the equilibrium condition.
The variation with respect to $n_+$ yields the following equation,
\begin{equation}
\frac{{\delta L}}{{\delta n_ +  }} = e_0\psi  - \mu _ +   - k_B T\left( { - \ln n_ +   - 1 + \ln \left( {N - n_ +   - n_ -   - n_{H^ +  }  - n_{OH^ -  } } \right) + 1} \right) = 0.
\label{eq:56}
\end{equation}
(Please refer to Supplementary Information information for the detailed derivation.)

We assume that the soft nanochannel connects two bulk reservoirs where the electrostatic potential and the number densities of different ions  are $\psi  = 0$  (please refer to \cite%
{Das_JAP_2015}) and $n_{i}=n_{ i ,\infty} \left(i=+,-,OH^-,H^+\right)$, respectively. Considering such facts, we obtain the following equation from Eq. (\ref{eq:56}).
\begin{equation}
\mu _ +   = k_B T\ln \frac{{n_{ +  ,\infty}}}{{N - n_{ + ,\infty }  - n_{ - ,\infty }  - n_{H^ +  ,\infty }  - n_{OH^ -  ,\infty } }}
\label{eq:6}
\end{equation}
Inserting the above equation in Eq. (\ref{eq:56})), we get the below equations
\begin{equation}
e_0\psi  + k_B T\ln \frac{{n_ +  \left( {N - n_{ + ,\infty }  - n_{ - ,\infty }  - n_{H^ +  ,\infty }  - n_{OH^ -  ,\infty } } \right)}}{{n_{ + ,\infty } \left( {N - n_ +   - n_ -   - n_{H^ +  }  - n_{OH^ -  } } \right)}} = 0,
\label{eq:7}
\end{equation}
After rearrangement of Eq. (\ref{eq:7}), the following equation is obtained 
\begin{equation}
n_ +   = n_{ + ,\infty } \frac{{N - n_ +   - n_ -   - n_{H^ +  }  - n_{OH^ -  } }}{{N - n_{ + ,\infty }  - n_{ - ,\infty }  - n_{H^ +  ,\infty }  - n_{OH^ -  ,\infty } }}\exp \left( { - \frac{{e_0\psi }}{{k_B T}}} \right).
\label{eq:8}
\end{equation}
Following such an argument as above-mentioned, $n_ -,  n_{OH^-}$ are obtained as follows
\begin{equation}
n_ -   = n_{ - ,\infty } \frac{{N - n_ +   - n_ -   - n_{H^ +  }  - n_{OH^ -  } }}{{N - n_{ + ,\infty }  - n_{ - ,\infty }  - n_{H^ +  ,\infty }  - n_{OH^ -  ,\infty } }}\exp \left( {\frac{{e_0\psi }}{{k_B T}}} \right),
\label{eq:10}
\end{equation}
\begin{equation}
n_{OH^ -  }  = n_{OH^ -  ,\infty } \frac{{N - n_ +   - n_ -   - n_{H^ +  }  - n_{OH^ -  } }}{{N - n_{ + ,\infty }  - n_{ - ,\infty }  - n_{H^ +  ,\infty }  - n_{OH^ -  ,\infty } }}\exp \left( {\frac{{e_0\psi }}{{k_B T}}} \right),
\label{eq:12}
\end{equation}
The variation Eq. (\ref {eq:215})  with respect to $n_{H^+}$ yields the following equation,
\begin{equation}
\frac{{\delta L}}{{\delta n_{H^+  }}} = e_0\psi  - \mu _{H^+}   - k_B T\left( { - \ln n_{H^+}   - 1 + \ln \left( {N - n_ +   - n_ -   - n_{H^ +  }  - n_{OH^ -  } } \right) + 1} \right)+e_0\varphi \left( y \right)\frac{{\gamma K'_a }}{{\left( {K'_a  + n_{H^ +  } } \right)^2 }}\psi = 0.
\label{eq:13}
\end{equation}
Following the argument in case of $n_{+}$ can provide $n_{H^+}$: 
\begin{equation}
n_{H^ +  }  = n_{H^ +  ,\infty } \frac{{N - n_ +   - n_ -   - n_{H^ +  }  - n_{OH^ -  } }}{{N - n_{ + ,\infty }  - n_{ - ,\infty }  - n_{H^ +  ,\infty }  - n_{OH^ -  ,\infty } }}\exp \left( { - \frac{{e_0\psi }}{{k_B T}}\left( {1 + \varphi \left( y \right)\frac{{\gamma K'_a }}{{\left( {K'_a  + n_{H^ +  } } \right)^2 }}} \right)} \right).
\label{eq:14}
\end{equation}
Dividing Eqs. (\ref{eq:8}, \ref{eq:12}, \ref{eq:14}) by Eq. (\ref{eq:10}) gives the below equations;
\begin{equation}
\frac{{n_ +  }}{{n_ -  }} = \frac{{n_{ + ,\infty } }}{{n_{ - ,\infty } }}\exp \left( { - \frac{{2e_0\psi }}{{k_B T}}} \right),
\label{eq:15}
\end{equation}

\begin{equation}
\frac{{n_{OH^ -  } }}{{n_ -  }} = \frac{{n_{OH^ -  ,\infty } }}{{n_{ - ,\infty } }},
\label{eq:16}
\end{equation}
\begin{equation}
\frac{{n_{H^ +  } }}{{n_ -  }} = \frac{{n_{H^ +  ,\infty } }}{{n_{ - ,\infty } }}\exp \left( { - \frac{{e_0\psi }}{{k_B T}}\left( {2 + \varphi \left( y \right)\frac{{\gamma K'_a }}{{\left( {K'_a  + n_{H^ +  } } \right)^2 }}} \right)} \right).
\label{eq:17}
\end{equation}
Substituting Eqs. (\ref{eq:15}, \ref{eq:16}, \ref{eq:17}) into Eq. (\ref{eq:10}) gives the following equations;
\begin{equation}
\small
n_ -   = n_{ - ,\infty } \frac{{{\rm{1 - }}n_ -  v_0 \left( {\frac{{n_{ + ,\infty } }}{{n_{ - ,\infty } }}\exp \left( { - \frac{{2e_0\psi }}{{k_B T}}} \right) + 1 + \frac{{n_{OH^ -  ,\infty } }}{{n_{ - ,\infty } }} + \frac{{n_{H^ +  ,\infty } }}{{n_{ - ,\infty } }}\exp \left( { - \frac{{e_0\psi }}{{k_B T}}\left( {2 + \varphi \left( y \right)\frac{{\gamma K'_a }}{{\left( {K'_a  + n_{H^ +  } } \right)^2 }}} \right)} \right)} \right)}}{{1 - \left( {n_{ + ,\infty }  + n_{ - ,\infty }  + n_{H^ +  }  + n_{OH^ -  } } \right)v_0 }}\exp \left( {\frac{{e_0\psi }}{{k_B T}}} \right),
\label{eq:18}
\end{equation}
After rearranging Eq. (\ref{eq:18}), the following equaion for $n_-$ is obtained,
\begin{equation}
n_ -   = \frac{{n_{ - ,\infty } \exp \left( {\frac{{e_0\psi }}{{k_B T}}} \right)}}{D},
\label{eq:19}
\end{equation}
where 
\begin{equation}
\begin{array}{l}
D=\left( {1 - n_{ + ,\infty } v_0  - n_{ - ,\infty } v_0  - n_{H^ +  ,\infty } v_0  - n_{OH^ -  ,\infty } v_0 } \right) + n_{ + ,\infty } v_0 \exp \left( { - \frac{{e_0\psi }}{{k_B T}}} \right) + n_{ - ,\infty } v_0 \exp \left( {\frac{{e_0\psi }}{{k_B T}}} \right) +\\
n_{OH^ -  ,\infty } v_0 \exp \left( {\frac{{e_0\psi }}{{k_B T}}} \right) + n_{H^ +  ,\infty } v_0 \exp \left( { - \frac{{e_0\psi }}{{k_B T}}\left( {1 + \varphi \left( y \right)\frac{{\gamma K'_a }}{{\left( {K'_a  + n_{H^ +  } } \right)^2 }}} \right)} \right).
\end{array}
\label{eq:195}
\end{equation}
Combining Eq. (\ref{eq:19}) and Eqs. (\ref{eq:15}, \ref{eq:16}, \ref{eq:17}) gives the number density equations for different species of ions;
\begin{equation}
n_ +   = \frac{{n_{ + ,\infty } \exp \left( { - \frac{{e_0\psi }}{{k_B T}}} \right)}}{D},
\label{eq:20}
\end{equation}
\begin{equation}
n_{OH^ -  }  = \frac{{n_{OH^ -  ,\infty } \exp \left( {\frac{{e_0\psi }}{{k_B T}}} \right)}}{D},
\label{eq:21}
\end{equation}
\begin{equation}
n_{H^ +  } = \frac{{n_{H^ +  ,\infty } \exp \left( { - \frac{{e_0\psi }}{{k_B T}}\left( {1 + \varphi \left( y \right)\frac{{\gamma K'_a }}{{\left( {K'_a  + n_{H^ +  } } \right)^2 }}} \right)} \right)}}{D}.
\label{eq:22}
\end{equation}
The expressions for the number densities of different ions are similar to those of \cite%
{Andelman_PRL_1997}, but should be implicitly determined.
 
On the other hand, minimizing Eq. (\ref{eq:215}) with respect to $\psi$  provides the equation for electrostatics.
\begin{equation}
\frac{{\delta L}}{{\delta \psi }} = 0 \Rightarrow \frac{{\partial f}}{{\partial \psi }} - \frac{d}{{dy}}\left( {\frac{{\partial f}}{{\partial \psi '}}} \right) \Rightarrow \frac{{d^2 \psi }}{{d^2 y^2 }} = \frac{{ - e_0\left( {n_ +   - n_ -  } \right) - e_0\left( {n_{H^ +  }  - n_{OH^ -  } } \right) + e_0\varphi \left( y \right)\frac{{K'_a \gamma }}{{K'_a  + n_{H^ +  } }}}}{{\varepsilon _0 \varepsilon _r }}.
\label{eq:23}
\end{equation}
In the region of $ - h + d \le y \le 0$, the equations for electrostatic potential and number densities of ions are reduced to those for \cite%
{Iglic_JPhysF_1996, Andelman_PRL_1997}.
\begin{equation}
\frac{{\delta L}}{{\delta \psi }} = 0 \Rightarrow \frac{{\partial f}}{{\partial \psi }} - \frac{d}{{dy}}\left( {\frac{{\partial f}}{{\partial \psi '}}} \right) \Rightarrow \frac{{d^2 \psi }}{{dy^2 }} = \frac{{ - e_0\left( {n_ +   - n_ -  } \right) - e_0\left( {n_{H^ +  }  - n_{OH^ -  } } \right)}}{{\varepsilon _0 \varepsilon _r }},
\label{eq:24}
\end{equation}
\begin{equation}
n_ -   = \frac{{n_{ - ,\infty } \exp \left( {\frac{{e_0\psi }}{{k_B T}}} \right)}}{D_0},
\label{eq:25}
\end{equation}
\begin{equation}
n_ +   = \frac{{n_{ + ,\infty } \exp \left( { - \frac{{e_0\psi }}{{k_B T}}} \right)}}{D_0},
\label{eq:26}
\end{equation}
\begin{equation}
n_{OH^ -  }  = \frac{{n_{OH^ -  ,\infty } \exp \left( {\frac{{e_0\psi }}{{k_B T}}} \right)}}{D_0},
\label{eq:27}
\end{equation}
\begin{equation}
n_{H^ +  }  = \frac{{n_{H^ +  ,\infty } \exp \left( { - \frac{{e_0\psi }}{{k_B T}}} \right)}}{D_0},
\label{eq:28}
\end{equation}
where
\begin{equation}
\begin{array}{l}
D_0=\left( {1 - n_{ + ,\infty } v_0  - n_{ - ,\infty } v_0  - n_{H^ +  ,\infty } v_0  - n_{OH^ -  ,\infty } v_0 } \right) + n_{ + ,\infty } v_0 \exp \left( { - \frac{{e_0\psi }}{{k_B T}}} \right) +\\
 n_{ - ,\infty } v_0 \exp \left( {\frac{{e_0\psi }}{{k_B T}}} \right) + n_{OH^ -  ,\infty } v_0 \exp \left( {\frac{{e_0\psi }}{{k_B T}}} \right) + n_{H^ +  ,\infty } v_0 \exp \left( { - \frac{{e_0\psi }}{{k_B T}}} \right).
\end{array}
\label{eq:255}
\end{equation}
(Please refer to Supplementary Information for the detailed derivation.)

We can easily justify that in order to ensure continuity in the value and its derivative of ion number densities, $ \varphi\left(y\right) $  must have the following properties;
\begin{equation}
\begin{array}{l}
\varphi _{y =  - h + d}  = 0,\\
\frac{{d\varphi }}{{dy}}_{y =  - h + d}  = 0,\\
\frac{{d\varphi }}{{dy}}_{y =  - h}  = 0.
\end{array}
\label{eq:26}
\end{equation}
These properties are the same as those in \cite%
{Das_RSC_2015} and finally such a function can be trivially obtained as follows; 
\begin{equation}
\varphi \left( y \right) = \beta \left( {\bar y + 1 - \bar d} \right)^2 \left( {\bar y + 1 + \frac{{\bar d}}{2}} \right),
\label{eq:27}
\end{equation}
where $\beta  = \frac{{4N_p a^3 h^3 }}{{\sigma d^4 }}$.
It should be noted that the cubic distribution of chargeable sites  is simplest among charge distributions which can remove  the discontinuities of physical quantities at the interface between polyelectrolyte layer and the interior of the channel.

As a consequence, non-uniform spatial distribution of chargeable sites remains unchanged irrespective whether non-zero value of ion size is considered, i.e. that of \cite%
{Das_RSC_2015}.


\end{document}